\documentclass{article}
\usepackage[T1]{fontenc}

\makeatletter

\newcommand{\LyX}{L\kern-.1667em\lower.25em\hbox{Y}\kern-.125emX\spacefactor1000}

\csname @addtoreset\endcsname{equation}{section}
\makeatother
\makeatother

\begin{document}

\newcommand{\hepth}[1]{hep-th/#1}

\newcommand{\intsum}{{\int\hspace{-3ex}\sum}}\begin{titlepage} 

\begin{flushright} {\small DFTT-7/99} \\ hep-th/9902181 \end{flushright} 

\vfill 

\begin{center} {

{\large On the Effective Potential of the ~\( Dp-\overline{Dp} \) system in type II theories.} 

\vskip 0.3cm

{\large {\sl }} 

\vskip 10.mm 

{\bf I. Pesando }\footnote{e-mail: ipesando@to.infn.it, pesando@alf.nbi.dk\\

Work supported by the European Commission TMR programme ERBFMRX-CT96-004}

\vskip4mm

{\small Dipartimento di Fisica Teorica , Universit\'a di Torino, via P. Giuria 1, I-10125 Torino, Istituto Nazionale di Fisica Nucleare (INFN) - sezione di Torino, Italy} 

}

\end{center} 

\vfill

\begin{center}{ \bf ABSTRACT}\end{center} 

\begin{quote}

We compute the effective potential of a system composed by a \( Dp \) brane and a separated \( \overline{Dp} \) brane at tree level in string theory. We show explicitly that the tachyon condenses and that the scalars which describe transverse fluctuations acquire~a vev proportional to the distance.


\vfill

\end{quote} 

\end{titlepage}

\section{Introduction.}

Since the beginning of the second string revolution most of the researches have
been devoted to the BPS spectrum since the BPS states can be reliably followed
from the weak to the strong regime. In the last period a mounting attention
has been dedicated to non BPS states which cannot decay because of some symmetries
or because of some topological obstruction (\cite{Sen-typeII},\cite{Sen-cond},\cite{Sen-altri},\cite{Ed},\cite{Yi}).
Many of these states have been constructed as bound states of brane-antibrane
systems with a topologically non trivial gauge configuration: it has then been
shown that these non BPS configurations are classified by the appropriate \( K \)-groups
of the spacetime where the branes are living. It has also been argued (\cite{Sen-cond})
that when the tachyon which is present in these systems condenses the vacuum
energy is equal and opposite to the brane tension, in such a way that the system
energy density is the same of the spacetime vacuum.

Our purpose is to perform an explicit tree level computation of the effective
potential on the brane belonging to a system made of a \( Dp \) and a parallel,
\emph{separated} \( \overline{Dp} \) in type II theory. A similar computation
for the case \( Dp-D(p+2) \) has already been performed in (\cite{GNS}) where
the computation turned out to be quite technical because they considered the
branes coincident and they should switch on a constant magnetic field in order
to be able to control the tachyon mass and then extract the exact quartic coupling
for the tachyon. We will use a different approach and we will consider separated
branes in such a way that the tachyon mass becomes zero at a certain distance
which we call the critical distance. It turns however out that this is not strictly
necessary and we are able to determine the quartic tachyonic potential for all
the distances in an interval around the critical distance, i.e. for all the
possible tachyonic mass such that \( \alpha 'm^{2}\rightarrow 0 \) as \( \alpha '\rightarrow 0 \),
in a unique way without any assumptions. In order to have a better check of
our computations we will consider a compact space in the brane directions and
we will switch a constant gauge field on in these directions.

The paper is organized as follows. In section 2 in order to fix our notations
we write the CFT of a bosonic string with total charge not zero propagating
on \( R\otimes T^{D-1} \) with a constant gauge field background. In section
3 we write the CFT for a system composed by a brane and a separated antibrane
in the bosonic string, in particular we write the vertex operator which creates
the tachyon present in such a configuration. In section 4 we perform the main
computation and we derive the coefficients entering the effective lagrangian.
Finally in section 5 we write explicitly the effective potential and we derive
our conclusions. In a series of appendices we carefully write down our conventions,
perform the QFT computations which are necessary to compare the string amplitudes
and write down the vertex operators which we use.

\section{The CFT of an open bosonic string on a toroidal space with constant Wilson
lines.}

We start considering the action for a open bosonic string in a background with
constant Wilson lines using the Polyakov formalism
\begin{equation}
\label{bos-act-con-wils-0}
S=-\frac{1}{4\pi \alpha '}\int d^{2}\xi \sqrt{-g\: }g^{\alpha \beta }\: \partial _{\alpha }X^{\mu }\partial _{\beta }X^{\nu }\: G_{\mu \nu }+e_{0}\: \int a_{\mu \: }dX^{\mu }+e_{\pi }\: \int a_{\mu \: }dX^{\mu }
\end{equation}
with \( G_{\mu \nu } \) the metric of the space \( R\otimes T^{D-1} \) and
\( e_{0}+e_{\pi }\neq 0 \).

If we limit to consider the diffeomorphisms which leave unchanged the range
of the spatial coordinate \( \xi ^{1}\equiv \sigma  \) as \( [0,\pi ] \) then
both \( \delta (\sigma )\: d^{2}\xi  \) and \( \delta (\pi -\sigma )\: d^{2}\xi  \)
are well defined and diffeomorphism invariant, so that the previous action (\ref{bos-act-con-wils-0})
can be rewritten as 
\begin{equation}
\label{bos-act-con-wils-1}
S=-\frac{1}{4\pi \alpha '}\int d^{2}\xi \left[ \sqrt{-g\: }g^{\alpha \beta }\: \partial _{\alpha }X^{\mu }\partial _{\beta }X^{\nu }\: G_{\mu \nu }+(-4\pi \alpha ')\left( e_{0}\: \delta (\sigma )+e_{\pi }\: \delta (\pi -\sigma )\right) a_{\mu \: }\partial _{0}X^{\mu }\right] 
\end{equation}
In this form it is evident that the Virasoro constraints are unchanged while
the canonical momentum gets a contribution from the Wilson lines.

In particular choosing the conformal gauge we get the usual solution for the
equations of motion:
\begin{equation}
\label{X-exp-in-modes}
X^{\mu }=x^{\mu }+2\alpha 'p^{\mu }\tau +i\sqrt{2\alpha '}\: \sum _{n\neq 0}\frac{1}{n}\alpha ^{\mu }_{n}e^{-in\tau }\cos n\sigma 
\end{equation}
which yields the usual hamiltonian
\begin{equation}
\label{L0}
H=L_{0}=\alpha 'p^{2}+\sum _{n=1}^{\infty }\alpha _{-n}\cdot \alpha _{n}
\end{equation}
but the canonical momentum 
\[
\Pi _{\mu }=\frac{\partial {\cal L}}{\partial \dot{X}^{\mu }}=\frac{1}{2\pi \alpha '}\dot{X}_{\mu }+\left( e_{0}\: \delta (\sigma )+e_{\pi }\: \delta (\pi -\sigma )\right) a_{\mu }\]
gets shifted\footnote{
We use the expansion \( \Pi _{\mu }=\frac{1}{\pi }\sum \pi _{n\mu }\cos n\sigma  \)
}. While the shift of the non zero modes has not effect, the shift on the zero
mode implies that the eigenvalues of \( p_{i} \) are not the usual \( \frac{n_{i}}{R_{(i)}} \)
but they are \( \frac{n_{i}}{R_{(i)}}+\pi \left( e_{0}+e_{\pi }\right) a_{i} \)
because the true conjugate momenta to the \( x^{\mu } \) variables are
\[
\pi _{0\mu }=p_{\mu }+\pi \left( e_{0}+e_{\pi }\right) a_{\mu },\; \; \; \; \left[ x^{\mu },\pi _{0\nu }\right] =i\: \delta ^{\mu }_{\nu }\]
whose eigenstates \( |k_{\mu }> \), \( \pi _{0\mu }|k>=k_{\mu }|k> \), have
the usual eigenvalues \( k_{i}=\frac{n_{i}}{R_{(i)}} \)because of the compactness
of the spatial directions.. This has the important consequence that the vacuum
of the theory \( |\Omega (a)> \)has not zero energy, i.e. it is not annihilated
by \( L_{0} \) generator (\ref{L0}). Moreover for some peculiar values of
\( \left( e_{0}+e_{\pi }\right) a_{i} \) the vacuum is degenerated, more explicitly
for every spatial direction \( i_{0} \) such that \( \pi \left( e_{0}+e_{\pi }\right) a_{i_{0}}=\frac{2l+1}{2R_{(i_{0})}} \)
with \( l\in Z \) we have a \( U(1) \) degeneration since all \( |\Omega (a),\theta >=\cos (\theta )|k=-l>+\sin (\theta )|k=-l-1> \),
where \( |k=l> \) is shorthand notation for the eigenstate of \( \pi _{0i_{0}} \)
with eigenvalue \( \frac{l}{R_{(i_{0})}} \), have the same vacuum energy.

The fact that the ground state \( |\Omega (a)> \) is not the \( sl(2) \) invariant
vacuum has another important effect, analogous to what happens in R sector of
the usual NSR theory and which can be exemplified by the following correspondence
between the tachyonic state and its vertex operator:
\begin{eqnarray}
|k,\Omega (a)>=e^{ik\cdot x}|\Omega (a)>\longleftrightarrow {\cal V}_{tach}(k;a;x)=:e^{i\left( k+\pi \left( e_{0}+e_{\pi }\right) a\right) \cdot X(x)}: &  & \label{V-tachione} \\
\alpha '\left( k+\pi \left( e_{0}+e_{\pi }\right) a\right) ^{2}=1 &  & \label{mass-shell-tachione} 
\end{eqnarray}
where the shift in the momentum in the exponent of the vertex operator is fundamental
for getting an operator with conformal dimension one, as one can easily verify
using the explicit expression for \( T(z) \) and OPEs which are unchanged w.r.t.
the case without Wilson line and the mass shell condition (\ref{mass-shell-tachione}).

\section{The CFT of two separated bosonic branes.}

We want now to write down the vertices describing open string states associated
with a system of two parallel separated branes or with a system of separated
brane-antibrane.

If we solve the equations of motion associated with the free bosonic action
with some directions \( i\in \left\{ 0,\ldots ,p\right\}  \) with Neumann boundary
condition and the remaining directions \( a\in \left\{ p+1,\ldots ,D-1\right\}  \)
satisfying Dirichlet boundary condition we would get that the D directions can
be expanded in modes as 
\begin{equation}
\label{X-D-exp-in-modes}
X^{a}=c^{a}+\frac{\Delta ^{a}}{\pi }\sigma +\sqrt{2\alpha '}\: \sum _{n\neq 0}\frac{1}{n}\alpha ^{a}_{n}e^{-in\tau }\sin n\sigma 
\end{equation}
 If we perform a canonical analysis of the system we would naively find that
\( c^{a} \) and \( \Delta ^{a} \) are pure numbers without corresponding conjugate
momenta. Nevertheless if we use Polchinski's trick (\cite{P}), i.e. we start
with \( Y^{a}=X^{a}_{L}(z)+X^{a}_{R}(\overline{z}) \) (\( z=e^{i(\tau +\sigma )} \),
\( \overline{z}=e^{i(\tau -\sigma )} \) ) with Neumann boundary condition and
with expansion in modes as in eq. (\ref{X-exp-in-modes}) and the we perform
a ``T-duality'' by letting 
\begin{equation}
\label{T-dual-open}
X^{a}_{R}(\overline{z})\rightarrow -X^{a}_{R}(\overline{z})
\end{equation}
we recover the wanted Dirichlet expansion (\ref{X-D-exp-in-modes}) with the
further information that 
\begin{equation}
\label{D-0-modes-bracket}
\frac{\Delta ^{a}}{\pi }\equiv 2\alpha 'p^{a}_{N},\: \: \: [y^{a},p^{b}_{N}]=i\eta ^{ab}
\end{equation}
which inform us that \( \frac{\Delta ^{a}}{\pi } \) is actually an operator
whose conjugate operator is given by \( \frac{1}{2\alpha '}y^{b} \). This result
is completely analogous to what happens in toroidal compactification in closed
string theory where the winding number turns up to be an operator only when
the T-dual formulation is considered.

Let us now exam the stress-energy tensor of this system; it is given by the
by now well known expression
\begin{equation}
\label{L0-D}
H=L_{0}=\alpha 'p^{i}p_{i}+\frac{1}{4\alpha '}\frac{\Delta ^{a}\Delta _{a}}{\pi ^{2}}+\sum _{n=1}^{\infty }\alpha _{-n}\cdot \alpha _{n}
\end{equation}
which can be trivially obtained from eq. (\ref{L0}) using the mapping (\ref{T-dual-open}).

The point with previous expression eq. (\ref{L0-D}) is that given a system
brane-(anti)brane separated by a distance \( \delta ^{a} \) the ground state
\( |\, \delta > \) is not \( sl(2) \) invariant, i.e. the energy vacuum is
not zero because of this it can be created from the \( sl(2) \) invariant vacuum
\( |\, 0> \) by acting with \( y^{a} \), explicitly
\begin{equation}
\label{D-vac-from-sl2-vac}
|\, \delta >=e^{i\frac{\delta }{\pi }\cdot \frac{y}{2\alpha '}}|\, 0>=\left( :e^{i\frac{\delta _{a}}{2\alpha '\pi }Y^{a}}:\right) (0)|\, 0>
\end{equation}
This is somewhat analogous to the relation between the R sector vacuum and the
NS vacuum.

As a consequence of this the mapping between states and vertices is a little
different from what naively we could have expected; let us consider for example
the tachyonic state \( |\, k_{i},\delta _{a};c^{a}> \) where the mass-shell
condition is \( \alpha 'k^{i}k_{i}=1-\frac{\delta ^{a}\delta _{a}}{4\alpha '\pi ^{2}} \)
, then the vertex operator which creates it from the \( sl(2) \) vacuum is
given by (a kind of ``Chan-Paton'' factor has to be understood in order to
keep track of the point where the string begins)
\begin{equation}
\label{bos-tach-oper}
{\cal V}_{tach}(k,\delta ;x)=:e^{ik_{i}X^{i}(x)}:\: :e^{i\frac{\delta _{a}}{2\alpha '\pi }Y^{a}(x)}:
\end{equation}
where it is important to notice the presence (not a priori obvious ) of the
coordinate expansion \( Y^{a} \) with Neumann boundary condition in the direction
in which we have imposed Dirichlet boundary condition; this is very important
since it assures that the vertex has the proper conformal dimension one. In
few words we can say that we use the old NSR tachyon vertex operator where we
reinterpret part of the momentum components as ``distance''.

\section{The potential of a brane-antibrane system in type II theory}

We have now all the ingredients necessary to write down the vertex operators
which describe the fields relevant to a low energy description of the system
made by the parallel couple \( Dp-\overline{Dp} \): 

\begin{itemize}
\item the \( U(1) \) field living on the brane \( A_{i} \) and the corresponding
living on the antibrane \( \overline{A}_{i} \) ;
\item the scalar fields describing the transverse oscillations \( \phi _{a} \) and
the corresponding living on the antibrane \( \overline{\phi }_{a} \) ;
\item the tachyonic field \( T \) describing the open string stretched between the
brane and the antibrane and its charge conjugate \( \overline{T} \) which describes
the open string with the opposite orientation.
\end{itemize}
Their explicit form is given in app. \ref{App_Vertex}: we do not write them
here because we do not need them for our explanation.

For these fields we can write down the most general quartic effective lagrangian
with at most two derivatives either as a lagrangian defined on the brane
\begin{eqnarray}
{\cal L}_{Dp}=-\frac{1}{4}F^{ij}F_{ij}-\frac{1}{2}\partial _{i}\phi _{a}\partial ^{i}\phi ^{a}-D_{i}T\overline{\: D^{i}T}-m^{2}T\overline{T} &  & \nonumber \\
-\nu _{a}\phi ^{a}T\overline{T}-\nu _{ab}\phi ^{a}\phi ^{b}T\overline{T}-\nu _{abcd}\: \phi ^{a}\phi ^{b}\phi ^{c}\phi ^{d}-\lambda \: \left( T\overline{T}\right) ^{2} &  & \label{action_QFT} 
\end{eqnarray}
(\( D_{i}=\partial _{i}-ie\: A_{i} \)) plus the analogous lagrangian for the
action defined on the antibrane when the brane and the antibrane can be distinguished
or as a unique lagrangian
\begin{eqnarray}
{\cal L}_{Dp-\overline{Dp}} & = & -\frac{1}{4}F^{ij}F_{ij}-\frac{1}{2}\partial _{i}\phi _{a}\partial ^{i}\phi ^{a}-\frac{1}{4}\overline{F}^{ij}\overline{F}_{ij}-\frac{1}{2}\partial _{i}\overline{\phi }_{a}\partial ^{i}\overline{\phi }^{a}\nonumber \\
 &  & -D_{i}T\overline{\: D^{i}T}-m^{2}T\overline{T}-2\lambda \left( T\overline{T}\right) ^{2}\nonumber \\
 &  & -\nu _{a}\phi ^{a}T\overline{T}-\overline{\nu }_{a}\overline{\phi }^{a}T\overline{T}\nonumber \\
 &  & -\nu _{ab}\phi ^{a}\phi ^{b}T\overline{T}-2\widetilde{\nu }_{ab}\phi ^{a}\overline{\phi }^{b}T\overline{T}-\overline{\nu }_{ab}\phi ^{a}\phi ^{b}T\overline{T}-\nu _{abcd}\phi ^{a}\phi ^{b}\phi ^{c}\phi ^{d}-\overline{\nu }_{abcd}\phi ^{a}\phi ^{b}\phi ^{c}\phi ^{d}\nonumber \label{action_QFT_totale} \\
 &  & \label{action_QFT_tot} 
\end{eqnarray}
(\( D_{i}=\partial _{i}-ie\: A_{i}-i\overline{e}\: \overline{A}_{i} \) ) when
the brane and antibrane are undistinguishable. In this latter case we have not
written the a priori possible potential terms mixing \( \phi  \) and \( \overline{\phi } \)
since it is not possible to draw a picture of the correspondig string amplitude.

In both cases we can immediately fix 
\[
\nu _{abcd}=\overline{\nu }_{abcd}=0\]
since the subset of this theory made by the \( U(1) \) gauge field and the
scalars can be obtained by a dimensional reduction of the \( {\cal N}=1 \)
\( D=10 \) super QED which is a free theory (moreover this can be checked by
an explicit string computation). From the knowledge of the open string spectrum
we can fix the mass of the tachyon when the branes are separated by a vector
\( \delta =(\delta ^{a}) \), normal to the brane and whose modulus is the distance,
to be given by 
\begin{equation}
\label{tachion_mass}
-m^{2}=\frac{1}{2\alpha '}-\left( \frac{\delta ^{a}}{2\pi \alpha '}\right) ^{2}
\end{equation}

Our purpose is now to determine all the remaining coefficients entering the
previous effective lagrangians (\ref{action_QFT},\ref{action_QFT_totale})
and then study the effective potential to see whether the tachyon condense and
whether other phenomena take place. To this purpose we first compute the 3-points
amplitudes, then the 4-points amplitudes and with the help of the factorization
property of the latter on different channels we determine the normalization
of the vertex operators. Finally taking the low energy limit we derive the wanted
coefficients as a function of the \( U(1) \) coupling constant \( e^{2} \).

In the following we partially use the notation of (\cite{GSW}), i.e. we denote
by \( W \) the ghosts and superghosts independent part of the corresponding
vertex operator \( {\cal V} \) in \( -1 \) picture and similarly for \( V \)
and the \( 0 \) picture, but we use the more conventional normal ordering for
the zero modes and we take the string worldsheet to be the upper complex plane.

\subsection{The 3-points amplitudes with one brane field.}

We start computing the \( \overline{T}TA_{i} \) amplitude in the \( F_{2} \)
formalism and 
\begin{eqnarray}
 &  & A_{\overline{T}TA_{i}}(k_{1};k_{2};k_{3},\zeta ;a_{i};\delta )=\nonumber \\
 & = & {\cal C}_{0}<0|W_{\overline{T}}(k_{1};a_{i};x=\infty )V_{T}(k_{2};a_{i};x=1)W_{A}(k_{3},\zeta ;x=0)|0>=\nonumber \\
 & = & (2\pi )^{p+1}\delta (\sum k)\: \frac{1}{2}2\alpha '{\cal C}_{0}{\cal N}^{2}_{T}{\cal N}_{A}V_{\bot }\: i\left[ \hat{k}_{2}-\widehat{\overline{k}}_{1}\right] \cdot \zeta \label{T-antiT-A} 
\end{eqnarray}
where \( \hat{k}_{2}=(k_{2}+\pi e_{0}a) \), \( \widehat{\overline{k}}_{1}=(k_{1}-\pi e_{0}a) \),
\( {\cal C}_{0} \) is the tree level amplitude normalization, \( {\cal N}_{T},{\cal N}_{A} \)
are the vertex operators normalizations discussed in app. \ref{App_Vertex}
following (\cite{DVLMR}), \( e_{0} \) is, as before in eq. (\ref{bos-act-con-wils-0}),
the charge of the open string which describes the (anti)tachyon, \( a_{i} \)
is the constant background \( U(1) \) field which we switch on on the brane
and \( \delta =\left( \delta ^{a}\right)  \) is the distance between the brane
and the antibrane.

We can now compare with the truncated amplitude computed from (\ref{action_QFT})
in app. \ref{App_QFT}:
\[
<\overline{T}(k_{1})T(k_{2})A_{i}(k_{3})>_{trunc}=-i(2\pi )^{p+1}\delta \left( \sum k\right) \: ie\left( \widehat{\overline{k}}_{1}-\hat{k}_{2}\right) _{i}\]
where \( \hat{k}_{i}=k_{i}-ea_{i} \) and \( \widehat{\overline{k}}_{i}=k_{i}+ea_{i} \).
This leads to the first two relations

\begin{eqnarray}
e & = & -\alpha '{\cal C}_{0}{\cal N}^{2}_{T}{\cal N}_{A}V_{\bot }\label{e_1} \\
e & = & -\pi e_{0}
\end{eqnarray}
which relates the \( U(1) \) coupling constant to the various normalization
factors and to the charge of the open string endpoint on the brane (see eq.
(\ref{bos-act-con-wils-0})).

We can now proceed further and compute the other non vanishing 3-point amplitude
which describes the coupling of the tachyon with the fluctuations transverse
to the brane:
\begin{eqnarray}
 &  & A_{\overline{T}T\phi }(k_{1};k_{2};k_{3},a;a_{i};\delta )=\nonumber \\
 & = & {\cal C}_{0}<0|W_{\overline{T}}(k_{1};a_{i};x=\infty )V_{T}(k_{2};a_{i};x=1)W_{\phi }(k_{3},a;x=0)|0>\nonumber \\
 & = & (2\pi )^{p+1}\delta (\sum k)\: 2\alpha '{\cal C}_{0}{\cal N}^{2}_{T}{\cal N}_{A}V_{\bot }\: i\frac{\delta ^{a}}{2\pi \alpha '}\label{T-antiT-fi} 
\end{eqnarray}
From the comparison of this amplitude with the corresponding one computed in
QFT as done in app. \ref{App_QFT}
\[
<\overline{T}(k_{1})T(k_{2})\phi _{a}(k_{3})>_{trunc}=-i(2\pi )^{p+1}\delta \left( \sum k\right) \: i\nu _{a}\]
we find that the parameter \( \nu _{a} \) is proportional to the distance between
the branes in the direction \( a \); more precisely we get
\begin{equation}
\label{nu_1}
\nu _{a}=2\alpha '{\cal C}_{0}{\cal N}^{2}_{T}{\cal N}_{A}V_{\bot }\frac{\delta ^{a}}{2\pi \alpha '}=-2e\frac{\delta ^{a}}{2\pi \alpha '}
\end{equation}
where we used eq. (\ref{e_1}) to write the last identity.

\subsection{The 3-points amplitudes with one antibrane field.}

We can now repeat all the computations of the previous subsection with the substitution
of the brane vertex operator with an antibrane vertex operator and with the
proper rearrangement of the tachyon and antitachyon vertex operators which is
necessary to be able to draw the corresponding graph. 

In particular we can compute the \( T\overline{T}\overline{A}_{i} \) amplitude
as done before and get
\begin{eqnarray}
 &  & A_{T\overline{T}\overline{A}_{i}}(k_{2};k_{1};k_{3},\zeta ;a_{i};\delta )=\nonumber \\
 & = & {\cal C}_{0}<0|W_{T}(k_{2};a_{i};x=\infty )V_{\overline{T}}(k_{1};a_{i};x=1)W_{\overline{A}}(k_{3},\zeta ;x=0)|0>=\nonumber \\
 & = & (2\pi )^{p+1}\delta (\sum k)\: \alpha '{\cal C}_{0}{\cal N}^{2}_{T}{\cal N}_{A}V_{\bot }\: (-i)\left[ \hat{k}_{2}-\widehat{\overline{k}}_{1}\right] \cdot \overline{\zeta }\label{T-antiT-Abar} 
\end{eqnarray}
and then compare with the truncated amplitude computed either from the analogous
for the antibrane of eq. (\ref{action_QFT}) or from eq. (\ref{action_QFT_totale}):
\[
<\overline{T}(k_{1})T(k_{2})\overline{A}_{i}(k_{3})>_{trunc}=-i(2\pi )^{p+1}\delta \left( \sum k\right) \: i\overline{e}\left( \widehat{\overline{k}}_{1}-\hat{k}_{2}\right) _{i}\]
 This leads to the expected relation
\begin{equation}
\label{ebar=-e}
\overline{e}=-e
\end{equation}
which relates the two \( U(1) \) coupling constants.

We can now compute the other non vanishing 3-point amplitude:
\begin{eqnarray}
 &  & A_{T\overline{T}\overline{\phi }}(k_{1};k_{2};k_{3},a;a_{i};\delta )=\nonumber \\
 & = & {\cal C}_{0}<0|W_{T}(k_{1};a_{i};x=\infty )V_{\overline{T}}(k_{2};a_{i};x=1)W_{\overline{\phi }}(k_{3},a;x=0)|0>\nonumber \\
 & = & (2\pi )^{p+1}\delta (\sum k)\: 2\alpha '{\cal C}_{0}{\cal N}^{2}_{T}{\cal N}_{A}V_{\bot }\: (-i)\frac{\delta ^{a}}{2\pi \alpha '}\label{T-antiT-fibar} 
\end{eqnarray}
which when compared with the QFT amplitude 
\[
<\overline{T}(k_{1})T(k_{2})\overline{\phi }_{a}(k_{3})>_{trunc}=-i(2\pi )^{p+1}\delta \left( \sum k\right) \: i\overline{\nu }_{a}\]
yields
\begin{equation}
\label{nubar_1}
\overline{\nu }_{a}=-\nu _{a}=+2e\frac{\delta ^{a}}{2\pi \alpha '}
\end{equation}
where we used eq. (\ref{e_1}) to write the last identity.

\subsection{The two tachyons - two antitachyons amplitude.}

We now want to compute the 2 tachyons - 2 antitachyons amplitude which is given
by the following expression in the \( F_{2} \) picture

\begin{eqnarray}
 &  & A_{\overline{T}T\overline{T}T}(k_{1};k_{2};k_{3};k_{4};a_{i})+A_{T\overline{T}T\overline{T}}(k_{2};k_{3};k_{4};k_{1};a_{i})=\nonumber \\
 & = & {\cal C}_{0}\int ^{1}_{0}dx<0|W_{\overline{T}}(k_{1};a_{i};z=\infty )V_{T}(k_{2};a_{i};z=1)V_{\overline{T}}(k_{3};a_{i};x)W_{T}(k_{4};a_{i};z=0)|0>\nonumber \\
 & + & {\cal C}_{0}\int ^{1}_{0}dx<0|W_{T}(k_{2};a_{i};z=\infty )V_{\overline{T}}(k_{3};a_{i};z=1)V_{T}(k_{4};a_{i};x)W_{\overline{T}}(k_{1};a_{i};z=0)|0>\nonumber \\
 & = & 2*(2\pi )^{p+1}\delta (\sum k)\: {\cal C}_{0}{\cal N}^{4}_{T}V_{\bot }\frac{\Gamma \left( 2\alpha 'K_{3}\cdot K_{4}+1\right) \Gamma \left( 2\alpha 'K_{2}\cdot K_{3}\right) }{\Gamma \left( 1+2\alpha 'K_{3}\cdot K_{4}+2\alpha 'K_{2}\cdot K_{3}\right) }\label{2T-2antiT} 
\end{eqnarray}
where we have introduced the following notations
\begin{eqnarray*}
K_{1,3} & = & \left( \widehat{\overline{k}}_{1,3},-\frac{\delta }{2\pi \alpha '}\right) =\left( k_{1,3}-\pi e_{0}a,-\frac{\delta }{2\pi \alpha '}\right) \\
K_{2,4} & = & \left( \widehat{k}_{2,4},\frac{\delta }{2\pi \alpha '}\right) =\left( k_{2,4}+\pi e_{0}a,\frac{\delta }{2\pi \alpha '}\right) 
\end{eqnarray*}
The introduction of the shifted momentum \( \hat{k} \) and \( \widehat{\overline{k}} \)
is justified by the mass shell condition (\ref{mass-shell-tachione}). By writing
this amplitude in this way it can be easily seen to be the old Lovelace amplitude
with a new interpretation, exactly as it happened with the bosonic tachyon vertex
operator (\ref{bos-tach-oper}). 

It is now worth discussing a subtlety concerning the previous amplitude (\ref{2T-2antiT})
which we wrote as the sum of the \( \overline{T}T\overline{T}T \) amplitude
and of the \( T\overline{T}T\overline{T} \) amplitude. At first site this latter
amplitude \( T\overline{T}T\overline{T} \) should not be here since it is the
ciclyc permutation ot the former: this is not true, this amplitude is different
from the first one considered since the intermediate states are antibrane states.
Another way of explaining why it has to be here is to notice that we have two
photons and each of them gives raise to its own \( s \) and \( t \) channel.

If we introduce the appropriate Mandelstam variables 
\begin{eqnarray}
s=-\left( \widehat{\overline{k}}_{1}+\hat{k}_{2}\right) ^{2} & t=-\left( \widehat{\overline{k}}_{1}+\hat{k}_{4}\right) ^{2} & u=-\left( \widehat{\overline{k}}_{1}+\widehat{\overline{k}}_{3}\right) ^{2}\nonumber \\
 & s+t+u=4m^{2} & \label{Mand_TTbarTTbar} 
\end{eqnarray}
then the low energy limit (\( |\alpha 's|\ll 1,|\alpha 't|\ll 1,|\alpha 'm^{2}|\ll 1 \))
of the amplitude (\ref{2T-2antiT}) can be computed as
\[
A_{\overline{T}T\overline{T}T}(k_{1};k_{2};k_{3};k_{4};a_{i})\sim (2\pi )^{p+1}\delta (\sum k)\: \frac{1}{\alpha '}{\cal C}_{0}{\cal N}^{4}_{T}V_{\bot }\frac{s+t}{st}\left( 1+\alpha '(s+t)\right) \]
It is important to notice that while one would think the factor \( \left( 1+\alpha '(s+t)\right)  \)
to be subleading, it is fundamental for getting both the proper \( s \) and
\( t \) channel factorization and the right quartic coefficient \( \lambda  \). 

Following the appendix of (\cite{DVLMR}) we get from the \( s\rightarrow 0 \)
channel factorization at any \( t \) on the brane (from the \( A_{\overline{T}T\overline{T}T} \)
only) the new relation 
\begin{equation}
\label{e2_piu_vincolo}
e^{2}=\frac{{\cal C}_{0}{\cal N}^{4}_{T}V_{\bot }}{2}\Rightarrow 2\alpha '^{2}{\cal C}_{0}{\cal N}^{2}_{A}V_{\bot }=1
\end{equation}
while all the other relations which arise in this and the other channel are
then trivially satisfied.

From the the 2 tachyons - 2 antitachyons QFT amplitude in the case of \emph{undistinguishable}
worldvolumes as it is can be easily deduced from what is carefully computed
in app. \ref{App_QFT} using eq. (\ref{action_QFT_totale})
\begin{eqnarray*}
 & <\overline{T}(k_{1})T(k_{2})\overline{T}(k_{3})T(k_{4})>_{trunc} & =\\
 & -i(2\pi )^{p+1}\delta \left( \sum k_{i}\right) \left\{ \left[ 4\lambda +\frac{e^{2}\left( t-u\right) +\nu _{a}\nu ^{a}}{s}+\frac{e^{2}\left( s-u\right) +\nu _{a}\nu ^{a}}{t}\right] \right.  & \\
 & +\left. \left[ 4\lambda +\frac{\overline{e}^{2}\left( t-u\right) +\overline{\nu }_{a}\overline{\nu }^{a}}{s}+\frac{\overline{e}^{2}\left( s-u\right) +\overline{\nu }_{a}\overline{\nu }^{a}}{t}\right] \right\}  & 
\end{eqnarray*}
and using the values obtained for \( e \) (\ref{e2_piu_vincolo}) and for \( \nu _{a} \)
(\ref{nu_1}) we can easily get that
\begin{equation}
\label{lambda}
\lambda =\frac{e^{2}}{2}
\end{equation}
We want to stress that this result is independent of the assumption that the
worldvolumes are undistinguishable since otherwise we should have considered
the two amplitudes \( A_{\overline{T}T\overline{T}T} \) and \( A_{T\overline{T}T\overline{T}} \)
separately and we would have obtained the same result.

The result is also independent of the values of the Mandelstam variables.

\subsection{The 4-points amplitude \protect\( \overline{T}T\phi \phi \protect \).}

Let us now turn our attention to the 4-points amplitude which we need: the \( \overline{T}T\phi \phi  \)
amplitude which is given by

\begin{eqnarray}
 &  & A_{\overline{T}T\phi \phi }(k_{1};k_{2};k_{3},a;k_{4},b;a_{i})=\nonumber \\
 &  & ={\cal C}_{0}\int ^{1}_{0}dx<0|W_{\overline{T}}(k_{1};a_{i};z=\infty )V_{T}(k_{2};a_{i};z=1)V_{\phi }(k_{3},a;x)W_{\phi }(k_{4},b;z=0)|0>+(a\leftrightarrow b)=\nonumber \\
 &  & =(2\pi )^{p+1}\delta (\sum k)\: \left( 2\alpha '\right) ^{2}{\cal C}_{0}{\cal N}^{2}_{T}{\cal N}^{2}_{A}V_{\bot }*\nonumber \\
 &  & \left[ \frac{\delta ^{a}}{2\pi \alpha '}\frac{\delta ^{b}}{2\pi \alpha '}\frac{\Gamma \left( 1+2\alpha 'k_{3}\cdot k_{4}\right) \Gamma \left( 2\alpha '\widehat{k}_{2}\cdot k_{3}\right) }{\Gamma \left( 1+2\alpha 'k_{3}\cdot k_{4}+2\alpha '\widehat{k}_{2}\cdot k_{3}\right) }-\frac{\delta ^{ab}}{2\alpha '}\frac{\Gamma \left( 2\alpha 'k_{3}\cdot k_{4}\right) \Gamma \left( 2\alpha '\widehat{k}_{2}\cdot k_{3}+1\right) }{\Gamma \left( 2\alpha 'k_{3}\cdot k_{4}+2\alpha '\widehat{k}_{2}\cdot k_{3}\right) }+(k_{3}\leftrightarrow k_{4})\right] \nonumber \\
 &  & \label{Tbar-T-fi-fi} 
\end{eqnarray}
When we define the Mandelstam variables for the \( T\overline{T}\rightarrow \phi \phi  \)
process 
\begin{eqnarray}
s=-\left( \widehat{\overline{k}}_{1}+\widehat{k}_{2}\right) ^{2} & t=-\left( \widehat{\overline{k}}_{1}+k_{4}\right) ^{2} & u=-\left( \widehat{\overline{k}}_{1}+k_{3}\right) ^{2}\nonumber \\
 & s+t+u=2m^{2} & \label{Mand-TbarTfifi} 
\end{eqnarray}
the low energy limit (\( |\alpha 's|\ll 1,|\alpha 't|\ll 1,|\alpha 'm^{2}|\ll 1 \))
is then given by
\begin{eqnarray*}
A_{\overline{T}T\phi \phi }(k_{1};k_{2};k_{3},a;k_{4},b;a_{i})\sim (2\pi )^{p+1}\delta (\sum k)\: \left( 2\alpha '\right) ^{2}{\cal C}_{0}{\cal N}^{2}_{T}{\cal N}^{2}_{A}V_{\bot } &  & \\
\left[ -\frac{1}{\alpha '}\frac{\delta ^{a}}{2\pi \alpha '}\frac{\delta ^{b}}{2\pi \alpha '}\left( \frac{1}{t-m^{2}}+\frac{1}{u-m^{2}}\right) -\frac{\delta ^{ab}}{2\alpha '}\right]  &  & 
\end{eqnarray*}

From the factorization in the \( t\sim m^{2} \), \( u\sim m^{2} \) channels
we get the condition 
\begin{equation}
\label{vincolo_su_NT}
\alpha '{\cal C}_{0}{\cal N}^{2}_{T}V_{\bot }=-1
\end{equation}
while comparing with the corresponding QFT amplitude 
\[
<\overline{T}(k_{1})T(k_{2})\phi _{a}(k_{3})\phi _{b}(k_{4})>_{trunc}=-i(2\pi )^{D}\delta \left( \sum k_{i}\right) \left( 2\nu _{ab}+\frac{\nu _{a}\nu _{b}}{u-m^{2}}+\frac{\nu _{a}\nu _{b}}{t-m^{2}}\right) \]
and using eq.s (\ref{nu_1},\ref{e2_piu_vincolo},\ref{vincolo_su_NT}) we find
the wanted quantity
\begin{equation}
\label{nu_1}
\nu _{ab}=e^{2}\delta _{ab}
\end{equation}

\subsection{The 4-points amplitude \protect\( T\overline{T}\overline{\phi }\overline{\phi }\protect \).}

The computations involved in this section are almost the same as the ones done
in the previous subsection; we have to compute the amplitude

\begin{eqnarray}
 &  & A_{T\overline{T}\overline{\phi }\overline{\phi }}(k_{1};k_{2};k_{3},a;k_{4},b;a_{i})=\nonumber \\
 &  & ={\cal C}_{0}\int ^{1}_{0}dx<0|W_{T}(k_{2};a_{i};z=\infty )V_{\overline{T}}(k_{1};a_{i};z=1)V_{\overline{\phi }}(k_{3},a;x)W_{\overline{\phi }}(k_{4},b;z=0)|0>+(a\leftrightarrow b)=\nonumber \\
 &  & =(2\pi )^{p+1}\delta (\sum k)\: \left( 2\alpha '\right) ^{2}{\cal C}_{0}{\cal N}^{2}_{T}{\cal N}^{2}_{A}V_{\bot }*\nonumber \\
 &  & \left[ \frac{\delta ^{a}}{2\pi \alpha '}\frac{\delta ^{b}}{2\pi \alpha '}\frac{\Gamma \left( 1+2\alpha 'k_{3}\cdot k_{4}\right) \Gamma \left( 2\alpha '\widehat{\overline{k}}_{1}\cdot k_{3}\right) }{\Gamma \left( 1+2\alpha 'k_{3}\cdot k_{4}+2\alpha '\widehat{\overline{k}}_{1}\cdot k_{3}\right) }-\frac{\delta ^{ab}}{2\alpha '}\frac{\Gamma \left( 2\alpha 'k_{3}\cdot k_{4}\right) \Gamma \left( 2\alpha '\widehat{\overline{k}}_{1}\cdot k_{3}+1\right) }{\Gamma \left( 2\alpha 'k_{3}\cdot k_{4}+2\alpha '\widehat{\overline{k}}_{1}\cdot k_{3}\right) }+(k_{3}\leftrightarrow k_{4})\right] \nonumber \\
 &  & \label{T-Tbar-fibar-fibar} 
\end{eqnarray}
write the low energy limit (\( |\alpha 's|\ll 1,|\alpha 't|\ll 1,|\alpha 'm^{2}|\ll 1 \))
using the Mandelstam variables (\ref{Mand-TbarTfifi})
\begin{eqnarray*}
A_{T\overline{T}\overline{\phi }\overline{\phi }}(k_{2};k_{1};k_{3},a;k_{4},b;a_{i})\sim (2\pi )^{p+1}\delta (\sum k)\: \left( 2\alpha '\right) ^{2}{\cal C}_{0}{\cal N}^{2}_{T}{\cal N}^{2}_{A}V_{\bot } &  & \\
\left[ -\frac{1}{\alpha '}\frac{\delta ^{a}}{2\pi \alpha '}\frac{\delta ^{b}}{2\pi \alpha '}\left( \frac{1}{t-m^{2}}+\frac{1}{u-m^{2}}\right) -\frac{\delta ^{ab}}{2\alpha '}\right]  &  & 
\end{eqnarray*}
 compare with the corresponding QFT amplitude 
\[
<\overline{T}(k_{1})T(k_{2})\overline{\phi }_{a}(k_{3})\overline{\phi }_{b}(k_{4})>_{trunc}=-i(2\pi )^{D}\delta \left( \sum k_{i}\right) \left( 2\overline{\nu }_{ab}+\frac{\overline{\nu }_{a}\overline{\nu }_{b}}{u-m^{2}}+\frac{\overline{\nu }_{a}\overline{\nu }_{b}}{t-m^{2}}\right) \]
and find the expected result
\begin{equation}
\label{nubar}
\overline{\nu }_{ab}=e^{2}\delta _{ab}
\end{equation}
We notice that factorization in the \( t \) and \( u \) channels does not
yield any new condition on the normalization coefficients.

\subsection{The 4-points amplitudes with one brane and one antibrane field.}

We now consider the \( T\phi \overline{T}\overline{\phi } \) amplitude that,
when we use the same momentum indexing as in eq. (\ref{Tbar-T-fi-fi}), is given
by

\begin{eqnarray*}
 &  & A_{T\phi \overline{T}\overline{\phi }}(k_{2};k_{3},a;k_{1};k_{4},b;a_{i})=\\
 &  & ={\cal C}_{0}\int ^{1}_{0}dx<0|W_{T}(k_{2};a_{i};z=\infty )V_{\phi }(k_{3},a;z=1)V_{\overline{T}}(k_{1};a_{1};x)W_{\overline{\phi }}(k_{4},b;z=0)|0>=\\
 &  & =(2\pi )^{p+1}\delta (\sum k)\: \left( 2\alpha '\right) ^{2}{\cal C}_{0}{\cal N}^{2}_{T}{\cal N}^{2}_{A}V_{\bot }*\\
 &  & \left[ -\frac{\delta ^{a}}{2\pi \alpha '}\frac{\delta ^{b}}{2\pi \alpha '}\frac{\Gamma \left( 2\alpha '\widehat{\overline{k}}_{1}\cdot k_{3}\right) \Gamma \left( 2\alpha '\widehat{\overline{k}}_{1}\cdot k_{4}\right) }{\Gamma \left( 2\alpha '\widehat{\overline{k}}_{1}\cdot k_{3}+2\alpha '\widehat{\overline{k}}_{1}\cdot k_{4}\right) }+\frac{\delta ^{ab}}{2\alpha '}\frac{\Gamma \left( 1+2\alpha '\widehat{\overline{k}}_{1}\cdot k_{3}\right) \Gamma \left( 1+2\alpha '\widehat{\overline{k}}_{1}\cdot k_{4}\right) }{\Gamma \left( 1+2\alpha '\widehat{\overline{k}}_{1}\cdot k_{3}+2\alpha '\widehat{\overline{k}}_{1}\cdot k_{4}\right) }\right] 
\end{eqnarray*}
Using the same Mandelstam variables (\ref{Mand-TbarTfifi}) as for the the \( \overline{T}T\phi \phi  \)
amplitude the low energy limit (\( |\alpha 'u|\ll 1,|\alpha 't|\ll 1,|\alpha 'm^{2}|\ll 1 \))
is then given by
\begin{eqnarray*}
A_{T\phi \overline{T}\overline{\phi }}(k_{2};k_{3},a;k_{1};k_{4},b;a_{i})\sim (2\pi )^{p+1}\delta (\sum k)\: \left( 2\alpha '\right) ^{2}{\cal C}_{0}{\cal N}^{2}_{T}{\cal N}^{2}_{A}V_{\bot } &  & \\
\left[ -\frac{1}{\alpha '}\frac{\delta ^{a}}{2\pi \alpha '}\frac{\delta ^{b}}{2\pi \alpha '}\left( \frac{1}{t-m^{2}}+\frac{1}{u-m^{2}}\right) +\frac{\delta ^{ab}}{2\alpha '}\right]  &  & 
\end{eqnarray*}

No further condition is obtained from the factorization in the \( t\sim m^{2} \),
\( u\sim m^{2} \) channels while the comparison with the corresponding QFT
amplitude 
\[
<\overline{T}(k_{1})T(k_{2})\phi _{a}(k_{3})\overline{\phi }_{b}(k_{4})>_{trunc}=-i(2\pi )^{D}\delta \left( \sum k_{i}\right) \left( 2\widetilde{\nu }_{ab}+\frac{\nu _{a}\overline{\nu }_{b}}{u-m^{2}}+\frac{\nu _{a}\overline{\nu }_{b}}{t-m^{2}}\right) \]
gives
\begin{equation}
\label{nutilde}
\widetilde{\nu }_{ab}=-e^{2}\delta _{ab}
\end{equation}

\section{The effective potentials and Conclusions.}

Before deriving the effective potentials a few words on the role of the GSO
projection in this derivation are now necessary: we have not mentioned the GSO
projector until now because it was not necessary for the computation of the
low energy action, would we have been interested in the higher derivatives corrections
then we should have inserted the proper GSO projector in the 4-points amplitudes
in order to prevent the propagation of unwanted states.

Let us now examine the meaning and the reliability of what we have done: for
distances near the critical one, i.e. such that \( \alpha 'm^{2}\rightarrow 0 \)
as \( \alpha '\rightarrow 0 \), the tachyon is almost massless and therefore
the low energy action must take it into account but now the critical distance
is \( \delta \sim \sqrt{\alpha '} \) which is of the order of the lowest distance
which can be probed by strings, therefore it cannot be asserted without doubts
that the two branes worldvolumes are distinguishable, hence we have to see the
consequences of assuming them to be distinguishable or to be undistinguishable.

We start assuming the worldvolumes to be separated and distinguishable in the
effective theory then, with all the constants computed, we can now explicitly
write down the potential as a function of the \( U(1) \) coupling constant,
of the distance \( \delta ^{a} \) and of the constant \( U(1) \) background
field \( a_{i} \) as
\begin{equation}
\label{eff-V}
V_{eff}=\left[ e^{2}a^{i}a_{i}-\frac{1}{2\alpha '}+\left( e\phi _{a}-\frac{\delta ^{a}}{2\pi \alpha '}\right) ^{2}\right] T\overline{T}+\frac{e^{2}}{2}\left( T\overline{T}\right) ^{2}
\end{equation}
This is the effective potential on the brane separated from the antibrane, the
effective potential on the antibrane can be obtained by \( e\rightarrow -e \)
.

This effective potential (\ref{eff-V}) shows that the tachyon condenses for
all the distances, thus breaking the \( U(1) \) gauge symmetry and that the
transverse fluctuations \( \phi  \) get an vev proportional to \( \delta  \).
In fact when the brane and the antibrane are parallel to the the axes and separated
only along the, let us say, the ninth direction we can shift \( \phi _{9}=\Phi +\frac{\delta }{2\pi \alpha '\: e} \)
and then we can write the mass part of the effective potential after the tachyon
condensation as
\begin{equation}
\label{mass-after-cond}
V_{mass}=\frac{1}{2\alpha 'e^{2}}\Phi ^{2}+\frac{1}{2\alpha '}A_{i}A^{i}
\end{equation}
which reveals that the only relevant low energy dof are the transverse fluctuations
in directions different from 9. This suggests an alternative way to the one
proposed in (\cite{Yi}) for explaining the fate of the gauge symmetry which
seems to survive on the worldvolume of the coincident brane-antibrane system
after the tachyon condensation when brane and antibrane annihilate: there is
actually no residual gauge symmetry to bother about. This can be justified by
noticing that for near critical distances both the \( U(1) \) on the brane
and the \( U(1) \) on the antibrane are broken, then we can try to argue that
this result is valid to all distances by noticing that we have apparently no
dependence on the distance on the tachyon condensation and that eq. (\ref{mass-after-cond})
can be applied for all distances. However\footnote{
We thank S.-J. Rey for pointing this out to us and for correspondance on the
subject.
}, the d.o.f. seem to be wrong because we have apparently only one d.o.f. associated
to the tachyon phase which we are using to give mass to two photons; one could
try to argue that the tachyon has a finite size even if it is massless and therefore
it can give nonlocal effects. This kind of argument can eventually be used to
explain the appearance of the amplitude \( T\overline{T}\rightarrow \phi \overline{\phi } \)
which from the brane worldvolume point of view is problematic since it would
be seen as \( T\overline{T}\rightarrow \phi  \) which violates the momentum
conservation. We want to point out that this kind of non locality is also present
in the usual parallel branes configuration, for example in the \( W^{+}W^{-}\rightarrow \gamma _{1}\gamma _{2} \),
where \( \gamma _{i} \) is the photon living on the i-th brane, but now it
happens at much higher energies, of the order of the Planck mass since this
is the \( W \) mass when worldvolumes are definitly distinguishable; we would
therefore expect to see some sign of this in the proper extension of Dirac-Born-Infeld
action since it can be interpreted as the effective action after the massive
mode have been integrated out. Nevertheless we think that this interpretation
is very problematic.

Let us now assume that the other possibility is true, namely that the branes
have an undistinguishable worldvolume even if the distance is almost stringy.
In this case it is not difficult to see that the effective potential of the
system is given by
\begin{equation}
\label{eff-V-tot}
V_{eff-tot}=\left[ e^{2}\left( a-\overline{a}\right) ^{i}\left( a-\overline{a}\right) _{i}-\frac{1}{2\alpha '}+\left( e\left( \phi -\overline{\phi }\right) _{a}-\frac{\delta ^{a}}{2\pi \alpha '}\right) ^{2}\right] T\overline{T}+2*\frac{e^{2}}{2}\left( T\overline{T}\right) ^{2}
\end{equation}

This effective potential (\ref{eff-V-tot}) shows that the tachyon condenses
for all the near to critical distances, thus breaking the non diagonal \( U(1)\subset U(1)\otimes U(1) \)
gauge symmetry since the tachyon couples now only to this gauge group as it
can be deduced from the effective action (\ref{action_QFT_totale}). In this
way we are left with the well known puzzle about the fate of the diagonal \( U(1) \)
but we solve the previous puzzle about the d.o.f..

It can also easily be seen that the transverse relative fluctuations \( \phi -\overline{\phi } \)
(which is the partner of \( A-\overline{A} \) in dimensional reduction) get
an vev proportional to \( \delta  \); let us again consider the same configuration
which led to eq. (\ref{mass-after-cond}), i.e. brane and antibrane parallel
to the the axes and separated only along the the ninth direction, then we can
shift \( \phi _{9}-\overline{\phi }_{9}=\Phi +\frac{\delta }{2\pi \alpha '\: e} \)
and write the mass part of the effective potential as
\begin{equation}
\label{mass-after-cond-tot}
V_{mass}=\frac{1}{2\alpha 'e^{2}}\Phi ^{2}+\frac{1}{2\alpha '}\left( A-\overline{A}\right) _{i}\left( A-\overline{A}\right) ^{i}
\end{equation}
 There is a nice interpretation for the shift \( \phi _{9}-\overline{\phi }_{9}=\Phi +\frac{\delta }{2\pi \alpha '\: e} \):
the brane and the antibrane tend to move closer, as one would naively expect
(\cite{Savv}), furthermore because of the mass which is acquired by the \( \Phi  \),
which decribes the relative oscillation, the system tends to remain stable once
it collapses. We can now ask ourselves what happens at substringy distances
where the tachyon is a real tachyon with an absolute mass of the order of the
Planck mass. The first observation is that we can probably still trust this
result since eq. (\ref{mass-after-cond-tot}) shows that both \( \Phi  \) and
\( A-\overline{A} \) have to be integrated out, their mass being independent
of \( \delta  \), and there is not any obvious mechanism which could make them
massless again. The second one is that no open string states can break the diagonal
\( U(1) \) since they are described by oriented strings which have opposite
charges w.r.t. the two different \( U(1) \)s.

As far as the vacuum energy density after the tachyon condensation we can easily
find 
\[
{\cal E}_{vacuum-pert}=-2*\frac{1}{8\alpha '^{2}\: e^{2}}\]
which is not the true non perturbative value which has been argued to be equal
to twice the brane tension (\cite{Sen-cond}) and which, in this units, is given
by 
\[
{\cal E}_{vacuum}=-2*\frac{1}{4\pi ^{2}\alpha '^{2}\: e^{2}}\]
as it can be deduce from the Dirac-Born-Infeld action
\[
S_{DBI}=\frac{1}{(2\pi \alpha ')^{2}e^{2}}\int d^{p+1}\xi \sqrt{-\det \left( \eta -2\pi \alpha 'e\: F\right) }\sim \int \frac{1}{(2\pi \alpha ')^{2}e^{2}}-\frac{1}{4}F^{2}\]
 This does not come as a surprise since there are not obvious symmetries which
could justify the coincidence of the perturbative and non-perturbative results.
This point of view is further supported by the fact that in (\cite{Sen-cond})
it was necessary to use non perturbative arguments to argue the equality of
the brane tension with the vacuum energy after the tachyon condensation.

It is nevertheless important to have found the Mexican hat potential at this
order since it would have been extremely difficult to justify the tachyon condensation
at higher order.

We can conclude by saying that we have found further evidence for the tachyon
condensation and shown how the brane-antibrane system becomes bound because
of tachyon condensation.

\bigskip{}
\textbf{Acknownledgments.}

\bigskip{}
We thank P. Di Vecchia, M.L. Frau, F. Gliozzi and A. Lerda for discussions.

\appendix

\section{Conventions.}

\begin{itemize}
\item Indexes: \( \mu ,\nu \in \left\{ 0,\ldots ,9\right\}  \),\( a,b\in \left\{ 0,\ldots ,p-1\right\}  \),
\( i,j\in \left\{ p,\ldots ,10\right\}  \)
\item Metric: \( G_{\mu \nu }=diag(-1,\underbrace{1,\ldots ,1}_{9}) \); \( g_{\alpha \beta }=diag(-1,1) \) 
\item Momentum eigenstates normalization in string amplitudes: \( <k|k'>=2\pi \: \delta (k-k')\rightarrow _{compact}2\pi R\: \delta _{k,k'} \)
\end{itemize}

\section{Scattering amplitudes in QFT on a toroidal space with constant Wilson lines.\label{App_QFT}}

We consider the following lagrangian which describes a complex tachyon with
quartic potential coupled to a \( U(1) \) gauge field and to \( n \) scalar
fields \( \phi  \) in \( D \) dimension:
\begin{eqnarray}
{\cal L}=-\frac{1}{4}F^{\mu \nu }F_{\mu \nu }-\frac{1}{2}\partial _{\mu }\phi _{\alpha }\partial ^{\mu }\phi ^{\alpha }-D_{\mu }T\overline{\: D^{\mu }T}-m^{2}T\overline{T} &  & \nonumber \\
-\nu _{\alpha }\phi ^{\alpha }T\overline{T}-\nu _{\alpha \beta }\phi ^{\alpha }\phi ^{\beta }T\overline{T}-\lambda \left( T\overline{T}\right) ^{2} &  & \label{app_action_QFT} 
\end{eqnarray}
where \( F_{\mu \nu }=\partial _{\mu }A_{\nu }-\partial _{\nu }A_{\mu } \),
\( D_{\mu }=\partial _{\mu }-ieA_{\mu } \), \( -m^{2}>0 \) and \( g_{\alpha \beta }=\delta _{\alpha \beta } \).

Now we want to compute some tree level scattering amplitudes, such as \( T\overline{T}\rightarrow T\overline{T} \),
on a spacetime \( R\otimes T^{D-1} \) in presence of a constant potential along
the compact directions \( A_{\mu }(x)=a_{\mu }+B_{\mu }(x) \) (\( a_{0}=0 \)
). To this purpose we define \( D_{(0)\mu }=\partial _{\mu }-iea_{\mu } \)
and we use as free gauge fixed lagrangian
\[
{\cal L}_{0}=-\frac{1}{4}\partial _{\mu }A_{\nu }\partial ^{\mu }A^{\nu }-\frac{1}{2}\partial _{\mu }\phi _{\alpha }\partial ^{\mu }\phi ^{\alpha }-D_{(0)\mu }T\overline{\: D_{(0)}^{\mu }T}-m^{2}T\overline{T}\]

We define the Fourier transform of a generic field \( \Phi (x) \) as

\[
\Phi (x)=\intsum \frac{dk}{(2\pi )^{\frac{D}{2}}}e^{ikx}\widetilde{\Phi }(k)\]

We can now apply the standard perturbation theory formula \( Z[J]=\int {\cal D}\Phi \exp \left( i\int {\cal L}+J\Phi \right) =\exp \left( -i\int d^{D}x\: V\left( -i\frac{\delta }{\delta J}\right) \right) Z_{0}[J] \)
where we have explicitly set \( \int J\Phi =\int d^{D}xT(x)\overline{J}(x)+\overline{T}(x)J(x)+B_{\mu }(x)J^{\mu }(x)+A_{\alpha }(x)J^{\alpha }(x) \)
and we obtained
\begin{equation}
\label{Z0}
Z_{0}[J]=\exp \left( i\intsum dk\frac{J^{\mu }(-k)J_{\mu }(k)}{2k^{2}}+\frac{J^{\alpha }(-k)J_{\alpha }(k)}{2k^{2}}+\frac{J(k)\overline{J}(-k)}{\hat{k}^{2}+m^{2}}\right) 
\end{equation}
where we have defined the following shifted momenta
\begin{eqnarray*}
\hat{k}_{\mu }=k_{\mu }-e_{\mu }\equiv k_{\mu }-ea_{\mu } &  & \\
\widehat{\overline{k}}_{\mu }=k_{\mu }+e_{\mu }\equiv k_{\mu }+ea_{\mu } &  & 
\end{eqnarray*}
 We easily get that the terms in \( Z[J] \) which contribute the \( T \) matrix
are
\[
Z[J]|_{J\overline{J}J_{\mu }}^{(tree)}=-i\intsum \prod _{i=1}^{3}\frac{dk_{i}}{(2\pi )^{\frac{D}{2}}}(2\pi )^{D}\delta \left( \sum k_{i}\right) \: \left( -ie\right) \left( \hat{k}_{1}^{\mu }-\widehat{\overline{k}}_{2}^{\mu }\right) \: \frac{iJ(k_{1})}{\hat{k}_{1}^{2}+m^{2}}\frac{i\overline{J}(k_{2})}{\widehat{\overline{k}}_{2}^{2}+m^{2}}\frac{iJ^{\mu }(k_{3})}{k_{3}^{2}}\]

\[
Z[J]|_{J\overline{J}J_{\alpha }}^{(tree)}=-i\intsum \prod _{i=1}^{3}\frac{dk_{i}}{(2\pi )^{\frac{D}{2}}}(2\pi )^{D}\delta \left( \sum k_{i}\right) \: i\nu _{\alpha }\: \frac{iJ(k_{1})}{\hat{k}_{1}^{2}+m^{2}}\frac{i\overline{J}(k_{2})}{\widehat{\overline{k}}_{2}^{2}+m^{2}}\frac{iJ^{\alpha }(k_{3})}{k_{3}^{2}}\]
and
\begin{eqnarray*}
Z[J]|_{J^{2}\overline{J}^{2}}^{(tree)}=-i\intsum \prod _{i=1}^{4}\frac{dk_{i}}{(2\pi )^{\frac{D}{2}}}(2\pi )^{D}\delta \left( \sum k_{i}\right) \frac{iJ(k_{1})}{\hat{k}_{1}^{2}+m^{2}}\frac{i\overline{J}(k_{2})}{\widehat{\overline{k}}_{2}^{2}+m^{2}}\frac{iJ(k_{3})}{\hat{k}_{3}^{2}+m^{2}}\frac{i\overline{J}(k_{4})}{\widehat{\overline{k}}_{4}^{2}+m^{2}} &  & \\
\left[ 4\lambda -e^{2}\frac{\left( \hat{k}_{2}-\widehat{\overline{k}}_{1}\right) \cdot \left( \hat{k}_{4}-\widehat{\overline{k}}_{3}\right) }{\left( k_{1}+k_{2}\right) ^{2}}-\frac{\nu _{\alpha }\nu ^{\alpha }}{\left( k_{1}+k_{2}\right) ^{2}}\right. \left. -e^{2}\frac{\left( \hat{k}_{4}-\widehat{\overline{k}}_{1}\right) \cdot \left( \hat{k}_{2}-\widehat{\overline{k}}_{3}\right) }{\left( k_{1}+k_{4}\right) ^{2}}-\frac{\nu _{\alpha }\nu ^{\alpha }}{\left( k_{1}+k_{4}\right) ^{2}}\right]  & 
\end{eqnarray*}
 
\[
Z[J]|_{J\overline{J}J^{2}_{\alpha }}^{(tree)}=-i\intsum \prod _{i=1}^{4}\frac{dk_{i}}{(2\pi )^{\frac{D}{2}}}(2\pi )^{D}\delta \left( \sum k_{i}\right) \frac{iJ(k_{1})}{\hat{k}_{1}^{2}+m^{2}}\frac{i\overline{J}(k_{2})}{\widehat{\overline{k}}_{2}^{2}+m^{2}}\frac{iJ^{\alpha }(k_{3})}{k_{3}^{2}}\frac{iJ^{\beta }(k_{4})}{k_{4}^{2}}\]
From these terms in the partition function we can compute the truncated Green
functions as 
\[
<\overline{T}(k_{1})T(k_{2})A_{\mu }(k_{3})>_{trunc}=-i(2\pi )^{D}\delta \left( \sum k_{i}\right) \: ie\left( \widehat{\overline{k}}_{1}^{\mu }-\hat{k}^{\mu }_{2}\right) \]

\[
<\overline{T}(k_{1})T(k_{2})\phi _{\alpha }(k_{3})>_{trunc}=-i(2\pi )^{D}\delta \left( \sum k_{i}\right) \: i\nu _{\alpha }\]

\begin{eqnarray*}
<\overline{T}(k_{1})T(k_{2})\phi _{\alpha }(k_{3})\phi _{\beta }(k_{4})>_{trunc}=-i(2\pi )^{D}\delta \left( \sum k_{i}\right)  &  & \\
\left( 2\nu _{\alpha \beta }-\frac{\nu _{\alpha }\nu _{\beta }}{\left( \widehat{\overline{k_{1}}}+k_{3}\right) ^{2}+m^{2}}-\frac{\nu _{\alpha }\nu _{\beta }}{\left( \widehat{\overline{k_{1}}}+k_{4}\right) ^{2}+m^{2}}\right)  &  & 
\end{eqnarray*}

\begin{eqnarray*}
<\overline{T}(k_{1})T(k_{2})\overline{T}(k_{3})T(k_{4})>_{trunc} & =-i(2\pi )^{D}\delta \left( \sum k_{i}\right)  & \\
\left[ 4\lambda -e^{2}\frac{\left( \hat{k}_{2}-\widehat{\overline{k}}_{1}\right) \cdot \left( \hat{k}_{4}-\widehat{\overline{k}}_{3}\right) }{\left( k_{1}+k_{2}\right) ^{2}}-\frac{\nu _{\alpha }\nu ^{\alpha }}{\left( k_{1}+k_{2}\right) ^{2}}\right.  & \left. -e^{2}\frac{\left( \hat{k}_{4}-\widehat{\overline{k}}_{1}\right) \cdot \left( \hat{k}_{2}-\widehat{\overline{k}}_{3}\right) }{\left( k_{1}+k_{4}\right) ^{2}}-\frac{\nu _{\alpha }\nu ^{\alpha }}{\left( k_{1}+k_{4}\right) ^{2}}\right] 
\end{eqnarray*}
where we have divided by \( \frac{1}{(2\pi )^{\frac{D}{2}}}\frac{1}{k^{2}} \)
for any \( \phi  \) and \( A \) external leg, \( \frac{1}{(2\pi )^{\frac{D}{2}}}\frac{1}{\widehat{\overline{k}}^{2}+m^{2}} \)
for any external \( \overline{T} \) leg and for \( \frac{1}{(2\pi )^{\frac{D}{2}}}\frac{1}{\hat{k}^{2}+m^{2}} \)
for any external \( T \) leg. 

The factors we have to use to find the truncated Green functions are derived
from the poles we can read in the 2-points Green function which can immediately
computed from (\ref{Z0}), hence we deduce that proper way to measure the momentum
for an antitachyon \( \overline{T}(k) \) is \( \widehat{\overline{k}} \) and
similarly for \( T(k) \) and \( \hat{k} \); moreover this is also supported
by the string mass shell condition (\ref{mass-shell-tachione}). As a consequence
of this observation we define the Mandelstam variables in the \( T\overline{T}\rightarrow T\overline{T} \)
amplitude as follows
\begin{eqnarray*}
s=-\left( \widehat{\overline{k}}_{1}+\hat{k}_{2}\right) ^{2} & t=-\left( \widehat{\overline{k}}_{1}+\hat{k}_{4}\right) ^{2} & u=-\left( \widehat{\overline{k}}_{1}+\widehat{\overline{k}}_{3}\right) ^{2}\\
 & s+t+u=4m^{2} & 
\end{eqnarray*}
so that we can write the amplitude as
\begin{eqnarray*}
<\overline{T}(k_{1})T(k_{2})\overline{T}(k_{3})T(k_{4})>_{trunc} & = & -i(2\pi )^{D}\delta \left( \sum k_{i}\right) \\
 &  & \left[ 4\lambda +\frac{e^{2}\left( t-u\right) +\nu _{\alpha }\nu ^{\alpha }}{s}+\frac{e^{2}\left( s-u\right) +\nu _{\alpha }\nu ^{\alpha }}{t}\right] 
\end{eqnarray*}

In a similar way we can introduce the proper Mandelstam variables for the amplitude
\( T\overline{T}\rightarrow \phi \phi  \) 
\begin{eqnarray*}
s=-\left( \widehat{\overline{k}}_{1}+\hat{k}_{2}\right) ^{2} & t=-\left( \widehat{\overline{k}}_{1}+k_{4}\right) ^{2} & u=-\left( \widehat{\overline{k}}_{1}+k_{3}\right) ^{2}\\
 & s+t+u=2m^{2} & 
\end{eqnarray*}
so that we can rewrite the truncated Green function as
\[
<\overline{T}(k_{1})T(k_{2})\phi _{\alpha }(k_{3})\phi _{\beta }(k_{4})>_{trunc}=-i(2\pi )^{D}\delta \left( \sum k_{i}\right) \left( 2\nu _{\alpha \beta }+\frac{\nu _{\alpha }\nu _{\beta }}{u-m^{2}}+\frac{\nu _{\alpha }\nu _{\beta }}{t-m^{2}}\right) \]

\section{String vertex operators for the system \protect\( Dp-\overline{Dp}\protect \)
with constant Wilson lines in the wordvolume of the brane.\label{App_Vertex}}

In writing the following vertex operators we denote with \( \psi ^{i} \) the
ws fermionic fields with indexes parallel to the brane, \( \Psi ^{a} \) the
ws fermionic fields with indexes perpendicular to the the brane and \( Y^{a} \)
the Neumann expansion corresponding to \( X^{a} \) which instead have Dirichlet
expansion.

In the text we use the same notations of (\cite{GSW}) and we denote by \( W_{T} \)
the part of \( {\cal V}_{T}^{(-1)} \) independent of ghosts and superghosts,
similarly for \( V_{T} \) and \( {\cal V}_{T}^{(0)} \) and all the other vertex
operators.

In all the next vertex operators a kind of ``Chan-Paton'' factor has to be
understood. Its role is to keep a record of the point in spacetime where the
string begins; without it there would be no information of this.

Tachyon vertex operators
\begin{eqnarray*}
{\cal V}_{T}^{(0)}(k,\delta ;a;x) & = & \sqrt{2\alpha '}{\cal N}_{T}\: c\: \left( \left( k+\pi e_{0}a\right) \cdot \psi +\frac{\delta }{2\pi \alpha '}\cdot \Psi \right) e^{i\left( k+\pi e_{0}a\right) \cdot X}e^{i\frac{\delta }{2\pi \alpha '}\cdot Y}\\
{\cal V}_{T}^{(-1)}(k,\delta ;a;x) & = & {\cal N}_{T}\: ce^{-\phi }\: e^{i\left( k+\pi e_{0}a\right) \cdot X}e^{i\frac{\delta }{2\pi \alpha '}\cdot Y}
\end{eqnarray*}
 We have inserted a \( \sqrt{2\alpha '} \) in the vertex in the \( 0 \) picture
because all the vertex operators must have the same engineering dimension in
spacetime.

Antitachyon vertex operators are obtained by \( \delta \rightarrow -\delta  \)
and \( e_{0}\rightarrow -e_{0} \).

\( U(1) \) gauge field vertex operators
\begin{eqnarray*}
{\cal V}^{(0)}_{A}(k,\zeta ;x) & = & i{\cal N}_{A}\: c\: \zeta \cdot \left( \dot{X}-2\alpha '\: \psi \: k\cdot \psi \right) e^{ik\cdot X}\\
{\cal V}^{(-1)}_{A}(k,\zeta ;x) & = & i\sqrt{2\alpha '}{\cal N}_{A}\: ce^{-\phi }\: \zeta \cdot \psi e^{ik\cdot X}
\end{eqnarray*}
We have again inserted a \( \sqrt{2\alpha '} \) in the vertex in the \( -1 \)
picture because all the vertex operators must have the same engineering dimension
in spacetime.

Transverse field \( \phi _{a} \) vertex operators
\begin{eqnarray*}
{\cal V}^{(0)}_{\phi }(k,a;x) & = & i{\cal N}_{A}\: c\: \left( X_{a}'-2\alpha '\: \Psi _{a}\: k\cdot \psi \right) e^{ik\cdot X}\\
{\cal V}^{(-1)}_{\phi }(k,\zeta ;x) & = & i\sqrt{2\alpha '}{\cal N}_{A}\: ce^{-\phi }\: \Psi _{a}e^{ik\cdot X}
\end{eqnarray*}
Here we have used the same normalization coefficient \( i{\cal N}_{A} \) as
for the \( U(1) \) gauge fields because the are in the same supermultiplet;
this can be checked directly from factorization. The presence of the \( \sigma  \)
derivative in the \( 0 \) picture vertex operator can be easily understood
in the old \( F_{2}-F_{1} \) formalism: infact starting from the vertex operator
\( W_{\phi }(\sigma =\tau =0) \) in the \( F_{1} \), a.k.a. \( -1 \) picture,
we compute the vertex in the \( F_{2} \) picture by \( V_{\phi }(\sigma =\tau =0)=[G_{r},W_{\phi }(\sigma =\tau =0)] \),
since the expressions for the \( G_{r} \) are the same as in the usual NSR
string then the \( V_{\phi }(\sigma =\tau =0) \) has the same expression of
the corresponding operator in the usual NSR theory in term of operators, but
now we have \( \dot{Y}(\sigma =0)=X'(\sigma =0) \). 

Using eq.s (\ref{e_1},\ref{e2_piu_vincolo},\ref{vincolo_su_NT}) we deduce
that the normalization factors are expressed through the \( U(1) \) coupling
constant \( e \) and \( \alpha ' \) as
\begin{eqnarray*}
{\cal N}_{A} & = & e\\
{\cal N}_{T} & = & i(2\alpha ')^{\frac{1}{2}}{\cal N}_{A}\\
{\cal C}_{0} & = & \frac{1}{2\alpha '^{2}V_{\bot }e^{2}}
\end{eqnarray*}


\begin{thebibliography}{}
\bibitem{P}J. Dai, R.G. Leigh and J.Polchinski, Mod. Phys. Lett. A4 (1989) 2073
\bibitem{BS}T. Banks and L. Susskind, Brane-Antibrane Forces, hepth 9511194
\bibitem{GNS}E. Gava, K.S. Narain and M.H. Sarmadi, On the Bound States of \( p- \) and
\( (p+2)- \)Branes, Nucl. Phys. B504 (1997) 214, hepth 9704006
\bibitem{Sen-typeII}A. Sen, Stable Non-BPS States in String Theory, JHEP 9806(1998)007, hepth 9803194\\
A. Sen, Stable Non-BPS Bound States of BPS D-branes, JHEP 9808(1998)010, hepth
9805019
\bibitem{Sen-cond}A. Sen, Tachyon Condensation on the Brane Antibrane System, JHEP 9808(1998)012,
hepth 9805170
\bibitem{Sen-altri}O. Bergman and M. R. Gaberdiel, Stable Non-BPS \( D \)-particles, Phys. Lett.
B441 (1998) 133, hepth 9806155\\
A. Sen, Type I D-particle and its Interaction, JHEP 9810(1998)021, hepth 9809111\\
A. Sen, BPS D-branes on Non-supersymmetric Cycles, JHEP 9812(1998)021, hepth
9812031\\
A. Sen, Descent Relations Among Bosonic D-branes, hepth 9902105
\bibitem{Savv}K. S. Savvidy, Brane Death from Born-Infeld String, hepth 9810163
\bibitem{Ed}G. Moore and R. Minasian, \( K \)-theory and Ramond-Ramond Charges, JHEP 9711(1997)002,
hepth 9710230\\
E. Witten, \( D \)-branes and \( K \)-theory, hepth 9810188 \\
P. Horava, Type \( IIA \) \( D \)-branes, \( K \)-theory, and Matrix Theory,
hepth 9812135\\
H. Garcìa-Compeàn, \( D \)-branes in Orbifold Singularities and Equivariant
\( K \)-theory, hepth 9812226\\
S. Gukov, \( K \)-theory, Reality and Orientifolds, hepth 9901042
\bibitem{Yi}P. Yi, Membranes from Five-Branes and Fundamental Strings from D\( p \) Branes,
hepth 9901159
\bibitem{DVLMR}P. Di Vecchia, A. Lerda, L. Magnea and R. Russo, String Techniques for the Calculation
of Renormalization Constants in Field Theory, Nucl.Phys.B469(1996)235, hepth
9601143
\bibitem{GSW}M.B. Green, J.H. Schwarz and E. Witten, Superstring Theory I, Cambridge University
Press
\end{thebibliography}
\end{document}